\documentstyle[12pt,epsfig]{article}
 
\newcommand{\agt}{\,\rlap{\lower 3.5 pt \hbox{$\mathchar \sim$}} \raise 1pt
 \hbox {$>$}\,}
\newcommand{\alt}{\,\rlap{\lower 3.5 pt \hbox{$\mathchar \sim$}} \raise 1pt
 \hbox {$<$}\,}

\catcode`@=11
\newcount\@tempcntc
\def\@citex[#1]#2{\if@filesw\immediate\write\@auxout{\string\citation{#2}}\fi
  \@tempcnta\z@\@tempcntb\m@ne\def\@citea{}\@cite{\@for\@citeb:=#2\do
    {\@ifundefined
       {b@\@citeb}{\@citeo\@tempcntb\m@ne\@citea\def\@citea{,}{\bf ?}\@warning
       {Citation `\@citeb' on page \thepage \space undefined}}%
    {\setbox\z@\hbox{\global\@tempcntc0\csname b@\@citeb\endcsname\relax}%
     \ifnum\@tempcntc=\z@ \@citeo\@tempcntb\m@ne
       \@citea\def\@citea{,}\hbox{\csname b@\@citeb\endcsname}%
     \else
      \advance\@tempcntb\@ne
      \ifnum\@tempcntb=\@tempcntc
      \else\advance\@tempcntb\m@ne\@citeo
      \@tempcnta\@tempcntc\@tempcntb\@tempcntc\fi\fi}}\@citeo}{#1}}
\def\@citeo{\ifnum\@tempcnta>\@tempcntb\else\@citea\def\@citea{,}%
  \ifnum\@tempcnta=\@tempcntb\the\@tempcnta\else
   {\advance\@tempcnta\@ne\ifnum\@tempcnta=\@tempcntb \else \def\@citea{--}\fi
    \advance\@tempcnta\m@ne\the\@tempcnta\@citea\the\@tempcntb}\fi\fi}
\catcode`@=12
\begin{document}

\title{\vskip-3cm{\baselineskip14pt
\centerline{\normalsize hep-ph/0101198\hfill}
\centerline{\normalsize January 2001\hfill}
}
\vskip1.5cm
Theoretical Aspects of Inclusive Light-Hadron Production
}
\author{{\sc Bernd A. Kniehl}\\
{\normalsize II. Institut f\"ur Theoretische Physik, Universit\"at Hamburg,}\\
{\normalsize Luruper Chaussee 149, 22761 Hamburg, Germany}}

\date{}

\maketitle

\thispagestyle{empty}

\begin{abstract}
We summarize the key results of a recent global analysis of inclusive
single-charged-hadron production in high-energy colliding-beam experiments.
In the framework of the parton model of quantum chromodynamics at 
next-to-leading order (NLO), fragmentation functions (FFs) for charged pions,
charged kaons, and (anti) protons are extracted from experimental data of
$e^+e^-$ annihilation at the $Z$-boson resonance and at centre-of-mass energy
$\sqrt s=29$~GeV.
This fit also yields a new NLO value for the strong-coupling constant, namely
$\alpha_s^{(5)}(M_Z)=0.1170\pm0.0073$.
The scaling violations encoded in the FFs through the Altarelli-Parisi
evolution equations are tested by confronting $e^+e^-$-annihilation data from
DESY DORIS, DESY PETRA, and CERN LEP2 with NLO predictions based on these FFs.
Comparisons of $p\bar p$ data from CERN S$p\bar p$S and the Fermilab Tevatron,
$\gamma p$ data from DESY HERA, and $\gamma\gamma$ data from LEP2 with the
corresponding NLO predictions allow us to test the universality of the FFs
predicted by the factorization theorem.
\medskip

\end{abstract}

%\newpage

\section{Introduction}

In the framework of the QCD-improved parton model, the inclusive production of
single hadrons is described by means of fragmentation functions (FFs),
$D_a^h(x,\mu^2)$.
The value of $D_a^h(x,\mu^2)$ corresponds to the probability for the parton
$a$ produced at short distance $1/\mu$ to form a jet that includes the hadron
$h$ carrying the fraction $x$ of the longitudinal momentum of $a$.
Unfortunately, it is not yet possible to calculate the FFs from first
principles, in particular for hadrons with masses smaller than or comparable
to the asymptotic scale parameter, $\Lambda$.
However, given their $x$ dependence at some energy scale $\mu$, the evolution
with $\mu$ may be computed perturbatively in QCD using the timelike 
Altarelli-Parisi equations \cite{gri}.
This allows us to test QCD quantitatively within one experiment observing
single hadrons at different values of centre-of-mass (CM) energy $\sqrt s$ (in
the case of $e^+e^-$ annihilation) or transverse momentum $p_T$ (in the case
of scattering).
Moreover, the factorization theorem guarantees that the $D_a^h(x,\mu^2)$
functions are independent of the process in which they have been determined,
and represent a universal property of $h$.
This enables us to make quantitative predictions for other types of
experiments as well.

After the pioneering leading-order (LO) analyses of pion, kaon, and
charmed-meson FFs in the late 1970s \cite{fie}, there had long been no
progress on the theoretical side of this field.
In the mid 1990s, next-to-leading-order (NLO) FF sets for $\pi^0$, $\pi^\pm$,
$K^\pm$, and $\eta$ mesons were constructed through fits to data of $e^+e^-$
annihilation, mostly generated with Monte Carlo event generators \cite{chi}.
In 1994/95, the author, in collaboration with Binnewies and Kramer, extracted
$\pi^{\pm}$ and $K^{\pm}$ FFs through fits to SLAC-PEP and partially
preliminary CERN-LEP1 data and thus determined the strong-coupling constant to
be $\alpha_s^{(5)}(M_Z)=0.118$ (0.122) at NLO (LO) \cite{bkk} (BKK).
However, these analyses suffered from the lack of specific data on the
fragmentation of tagged quarks and gluons to $\pi^\pm$, $K^\pm$, and
$p/\bar p$ hadrons.

During the last five years, the experiments at LEP1 and SLAC SLC have provided
us with a wealth of high-precision information on how partons fragment into
low-mass charged hadrons, so as to cure this problem.
The data partly comes as light-, $c$-, and $b$-quark-enriched samples without
\cite{Al,A,D} or with identified final-state hadrons ($\pi^\pm$, $K^\pm$, and
$p/\bar p$) \cite{A1,D1,S} or as gluon-tagged three-jet samples without hadron
identification \cite{Ag,Og,Dg}.
Motivated by this new situation, the author, together with Kramer and 
P\"otter, recently updated, refined, and extended the BKK \cite{bkk} analysis
by generating new LO and NLO sets of $\pi^\pm$, $K^\pm$, and $p/\bar p$ FFs
\cite{kkp}.\footnote{%
A FORTRAN subroutine which returns the values of the $D_a^h(x,\mu^2)$
functions for given values of $x$ and $\mu^2$ may be downloaded from the URL
{\tt http://www.desy.de/\~{}poetter/kkp.html} or obtained upon request from
the authors.}
By also including in our fits $\pi^\pm$, $K^\pm$, and $p/\bar p$ data (without
flavour separation) from PEP \cite{T}, with CM energy $\sqrt s=29$~GeV, we
obtained a handle on the scaling violations in the FFs, which enabled us to
determine the strong-coupling constant.
We found $\alpha_s^{(5)}(M_Z)=0.1181\pm0.0085$ at LO and
$\alpha_s^{(5)}(M_Z)=0.1170\pm0.0073$ at NLO \cite{kkp1}.
These results are in perfect agreement with what the Particle Data Group
currently quotes as the world average, $\alpha_s^{(5)}(M_Z)=0.1181\pm0.002$
\cite{pdg}.
Our strategy was to only include in our fits LEP1 and SLC data with both
flavour separation and hadron identification \cite{A1,D1,S}, gluon-tagged
three-jet samples with a fixed gluon-jet energy \cite{Ag,Og}, and the
$\pi^\pm$, $K^\pm$, and $p/\bar p$ data sets from the pre-LEP1/SLC era
with the highest statistics and the finest binning in $x$ \cite{T}.
Other data served us for cross checks.
In particular, we probed the scaling violations in the FFs through comparisons
with $\pi^\pm$, $K^\pm$, and $p/\bar p$ data from DESY DORIS and DESY PETRA,
with CM energies between 5.4 and 34~GeV \cite{low}.
Furthermore, we tested the gluon FF, which enters the unpolarized cross
section only at NLO, by comparing our predictions for the longitudinal cross
section, where it already enters at LO, with available data \cite{Al,Dl}.
Finally, we directly compared our gluon FF with the one recently measured by
DELPHI in three-jet production with gluon identification as a function of $x$
at various energy scales $\mu$ \cite{Dg}.
All these comparisons led to rather encouraging results.
We also verified that our FFs satisfy reasonably well the momentum sum
rules, which we did not impose as constraints on our fits.

Very recently, we extended our previous tests of scaling violations \cite{kkp}
to higher energy scales by confronting new data on inclusive charged-hadron
production in $e^+e^-$ annihilation from LEP2 \cite{D2}, with $\sqrt s$
ranging from 133~GeV up to 189~GeV, with NLO predictions based on our FFs
\cite{kkp2}.
Furthermore, we quantitatively checked the universality of our FFs by making
comparisons with essentially all available high-statistics data on inclusive
charged-hadron production in colliding-beam experiments \cite{kkp2}.
This includes $p\bar p$ data from the UA1 and UA2 Collaborations \cite{UA1} at
CERN S$p\bar p$S and from the CDF Collaboration \cite{CDF} at the Fermilab
Tevatron, $\gamma p$ data from the H1 and ZEUS Collaborations \cite{H1} at
DESY HERA, and $\gamma\gamma$ data from the OPAL Collaboration \cite{Ogg} at
LEP2.

In 2000, alternative sets of NLO FFs for $\pi^\pm$, $K^\pm$ \cite{kre}, and
charged hadrons \cite{kre,bou} have become available.
They are based on different collections of experimental data and on additional
theoretical assumptions.
In Ref.~\cite{kre}, power laws were assumed to implement a hierarchy among the
valence- and sea-quark FFs.
In Ref.~\cite{bou}, the renormalization and factorization scales were 
identified and adjusted according to the principle of minimal sensitivity
\cite{ste}.
In order to estimate the present systematic uncertainties in the FFs, we
compare NLO predictions for inclusive charged-hadron production consistently
evaluated with the three new-generation FF sets \cite{kkp,kre,bou}.

In this contribution, we summarize the key results obtained in 
Refs.~\cite{kkp,kkp1,kkp2}.
In Section~\ref{sec:two}, we present some details of our global fits and
assess the quality of the resulting FFs.
In Section~\ref{sec:three}, we discuss the determination of
$\alpha_s^{(5)}(M_Z)$ from the scaling violations in the FFs and compare our
NLO result with those of different determinations.
In Section~\ref{sec:four}, we present comparisons of our NLO predictions for
inclusive charged-hadron production with $e^+e^-$ data from LEP2 \cite{D2},
$p\bar p$ data from S$p\bar p$S \cite{UA1} and the Tevatron \cite{CDF},
$\gamma p$ data from HERA \cite{H1}, and $\gamma\gamma$ data from LEP2
\cite{Ogg}.
Our conclusions are summarized in Section~\ref{sec:five}.

\section{Determination of the FFs
\label{sec:two}}

The NLO formalism for extracting FFs from measurements of the cross section
$d\sigma/dx$ of inclusive hadron production in $e^+e^-$ annihilation was
comprehensively described in Ref.~\cite{bkk}.
We work in the $\overline{\mathrm{MS}}$ renormalization and factorization
scheme and choose the renormalization scale $\mu$ and the factorization scale
$M_f$ to be $\mu=M_f=\xi\sqrt s$, except for gluon-tagged three-jet events,
where we put $\mu=M_f=2\xi E_{\mathrm{jet}}$, with $E_{\mathrm{jet}}$ being
the gluon jet energy in the CM frame.
Here, the dimensionless parameter $\xi$ is introduced to determine the
theoretical uncertainty in $\alpha_s^{(5)}(M_Z)$ from scale variations.
As usual, we allow for variations between $\xi=1/2$ and 2 around the default
value 1.
For the actual fitting procedure, we use $x$ bins in the interval
$0.1\le x\le 1$ and integrate the theoretical functions over the bin widths as
is done in the experimental analyses.
The restriction at small $x$ is introduced to exclude events in the region
where mass effects and nonperturbative intrinsic-transverse-momentum effects
are important and the underlying formalism is insufficient.
On the other hand, our analysis should be rather insensitive to
nonperturbative effects at $x$ values close to 1, since the experimental 
errors are very large there.
We parameterize the $x$ dependence of the FFs at the starting scale $\mu_0$ as  
$D_a^h(x,\mu_0^2)=Nx^{\alpha}(1-x)^{\beta}$.
We treat $N$, $\alpha$, and $\beta$ as independent fit parameters.
In addition, we take the asymptotic scale parameter
$\Lambda_{\overline{\mathrm{MS}}}^{(5)}$, appropriate for five quark flavors,
as a free parameter.
Thus, we have a total of 46 independent fit parameters.
The quality of the fit is measured in terms of the $\chi^2$ value per degree
of freedom, $\chi^2_{\mathrm{DF}}$, for all selected data points.
Using a multidimensional minimization algorithm \cite{jam}, we search this
46-dimensional parameter space for the point at which the deviation of the
theoretical prediction from the data becomes minimal.

\begin{table}[t]
\caption{CM energies, types of data, and $\chi^2_{\mathrm{DF}}$ values
obtained at LO and NLO for the various data samples.}
\label{tab:one}
\begin{tabular}{c|l|ll}
 $\sqrt{s}$ [GeV] & Data type & 
 \multicolumn{2}{c}{\makebox[4.5cm][c]{$\chi^2_{\mathrm{DF}}$ in NLO (LO)}} \\
 \hline
29.0 & $\sigma^\pi$~(all) & 0.64 (0.71) \cite{T} & \\
     & $\sigma^K$~(all) & 1.86 (1.40) \cite{T} & \\
     & $\sigma^p$~(all) & 0.79 (0.70) \cite{T} & \\ \hline
91.2 & $\sigma^h$~(all) & 1.28 (1.40) \cite{D1} & 1.32 (1.44) \cite{S} \\
     & $\sigma^h$~(uds) & 0.20 (0.20) \cite{D1} & \\
     & $\sigma^h$~(b)   & 0.43 (0.41) \cite{D1} & \\ \hline
     & $\sigma^\pi$~(all) & 1.28 (1.65) \cite{A1} & \\
     &                    & 0.58 (0.60) \cite{D1} & 3.09 (3.13) \cite{S} \\
     & $\sigma^\pi$~(uds) & 0.72 (0.73) \cite{D1} & 1.87 (2.17) \cite{S} \\
     & $\sigma^\pi$~(c) & & 1.36 (1.16) \cite{S} \\
     & $\sigma^\pi$~(b) & 0.57 (0.58) \cite{D1} & 1.00 (0.99) \cite{S} \\
\hline
     & $\sigma^K$~(all) & 0.30 (0.32) \cite{A1} & \\
     &                  & 0.86 (0.79) \cite{D1} & 0.44 (0.45) \cite{S} \\
     & $\sigma^K$~(uds) & 0.53 (0.60) \cite{D1} & 0.65 (0.64) \cite{S} \\
     & $\sigma^K$~(c) & & 2.11 (1.90) \cite{S} \\
     & $\sigma^K$~(b) & 0.14 (0.14) \cite{D1} & 1.21 (1.23) \cite{S} \\ \hline
     & $\sigma^p$~(all) & 0.93 (0.80) \cite{A1} & \\
     &                  & 0.09 (0.06) \cite{D1} & 0.79 (0.70) \cite{S} \\
     & $\sigma^p$~(uds) & 0.11 (0.14) \cite{D1} & 1.29 (1.28) \cite{S} \\
     & $\sigma^p$~(c)   & & 0.92 (0.89) \cite{S} \\
     & $\sigma^p$~(b)   & 0.56 (0.62) \cite{D1} & 0.97 (0.89) \cite{S} \\
\hline
$E_{\mathrm{jet}}$ [GeV] & & & \\ \hline
26.2 & $D_g^h$ & 1.19 (1.18) \cite{Ag} & \\
40.1 & $D_g^h$ & 1.03 (0.90) \cite{Og} & \\
\end{tabular}
\end{table}

\begin{figure}[ht]
\begin{center}
\begin{tabular}{ll}
\parbox{8cm}{\epsfig{figure=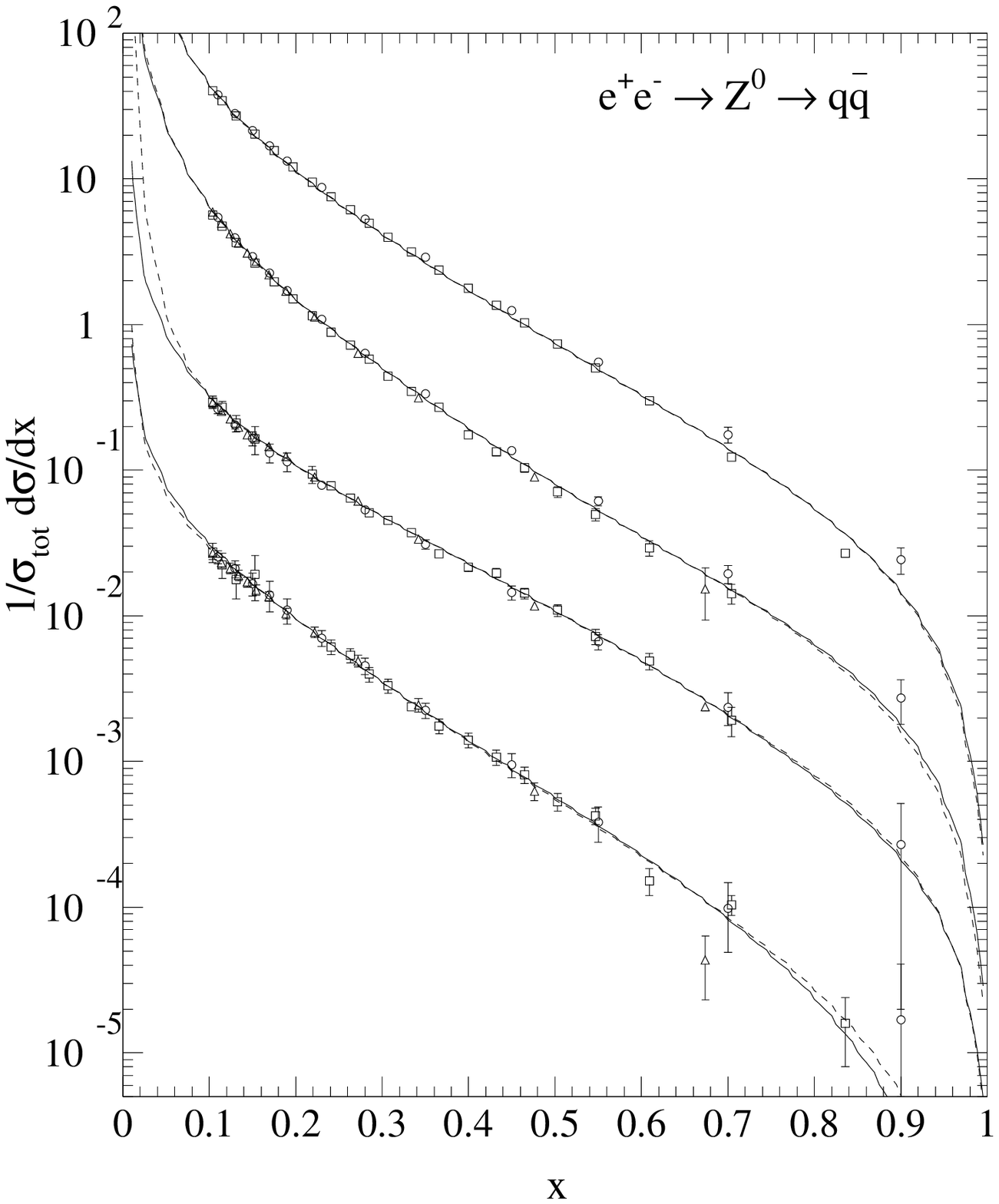,width=8cm}} &
\parbox{8cm}{\epsfig{figure=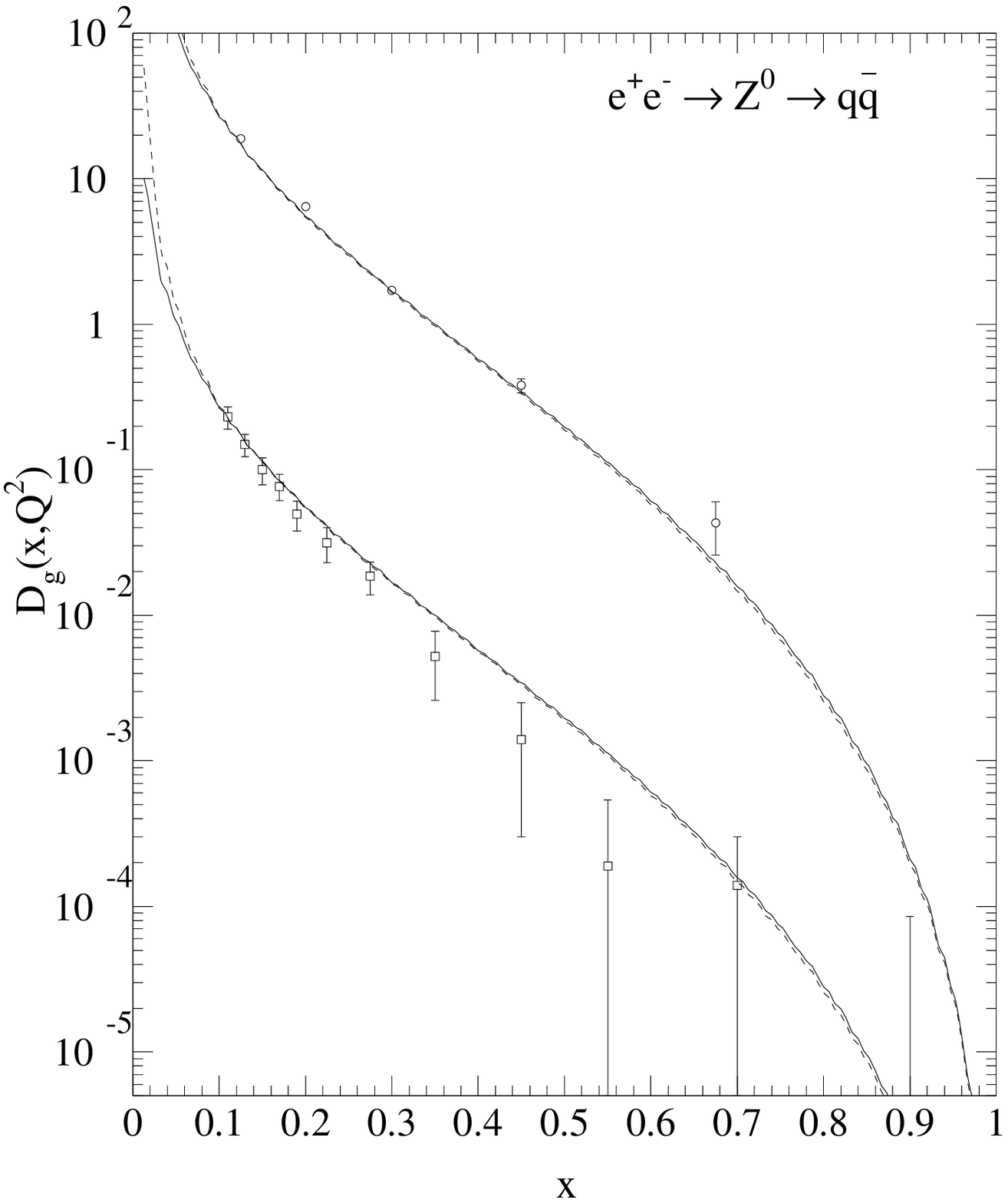,width=8cm}} \\
(a) & (b) \\
\end{tabular}
\caption{(a) Comparison of data on inclusive charged-hadron production at
$\protect\sqrt{s}=91.2$~GeV from ALEPH \protect\cite{A1} (triangles), DELPHI 
\protect\cite{D1} (circles), and SLD \protect\cite{S} (squares) with our LO
(dashed lines) and NLO (solid lines) fit results.
The upmost, second, third, and lowest curves refer to charged hadrons,
$\pi^\pm$, $K^\pm$, and $p/\bar{p}$, respectively.
(b) Comparison of three-jet data on the gluon FF from ALEPH \protect\cite{Ag}
with $E_{\mathrm{jet}}=26.2$~GeV (upper curves) and from OPAL
\protect\cite{Og} with $E_{\mathrm{jet}}=40.1$~GeV (lower curves) with our LO
(dashed lines) and NLO (solid lines) fit results.}
\label{fig:one}
\end{center}
\end{figure}

The $\chi^2_{\mathrm{DF}}$ values achieved for the various data sets used in
our LO and NLO fits may be seen from Table~\ref{tab:one}.
Most of the $\chi^2_{\mathrm{DF}}$ values lie around unity or below,
indicating that the fitted FFs describe all data sets within their respective
errors. 
In general, the $\chi^2_{\mathrm{DF}}$ values come out slightly in favor for
the DELPHI \cite{D1} data.
The overall goodness of the NLO (LO) fit is given by
$\chi^2_{\mathrm{DF}}=0.98$ (0.97).
The goodness of our fit may also be judged from Figs.~\ref{fig:one}(a) and
\ref{fig:one}(b), where our LO and NLO fit results are compared with the ALEPH
\cite{A1,Ag}, DELPHI \cite{D1}, OPAL \cite{Og}, and SLD \cite{S} data.
In Fig.~\ref{fig:one}(a), we study the differential cross section
$(1/\sigma_{\mathrm{tot}})d\sigma^h/dx$ for $\pi^\pm$, $K^\pm$, $p/\bar p$,
and unidentified charged hadrons at $\sqrt{s}=91.2$~GeV, normalized to the
total hadronic cross section $\sigma_{\mathrm{tot}}$, as a function of the
scaled momentum $x=2p_h/\sqrt s$.
As in Refs.~\cite{D1,S}, we assume that the sum of the $\pi^\pm$, $K^\pm$, and
$p/\bar p$ data exhaust the full charged-hadron data.
We observe that, in all cases, the various data are mutually consistent with
each other and are nicely described by the LO and NLO fits, which is also
reflected in the relatively small $\chi^2_{\mathrm{DF}}$ values given in 
Table~\ref{tab:one}.
The LO and NLO fits are almost indistinguishable in those regions of $x$,
where the data have small errors.
At large $x$, where the statistical errors are large, the LO and NLO results 
sometimes moderately deviate from each other.
In Fig.~\ref{fig:one}(b), we compare the ALEPH \cite{Ag} and OPAL \cite{Og}
measurements of the gluon FF in gluon-tagged charged-hadron production, with
$E_{\mathrm{jet}}=26.2$ and 40.1~GeV, respectively, with our LO and NLO fit
results.
The data are nicely fitted, with $\chi^2_{\mathrm{DF}}$ values of order unity,
as may be seen from Table~\ref{tab:one}.
By the same token, this implies that these data sets \cite{Ag,Og} are mutually
consistent.

\boldmath
\section{Determination of $\alpha_s^{(5)}(M_Z)$
\label{sec:three}}
\unboldmath

Since we included in our fits high-quality data from two very different
energies, namely 29 and 91.2~GeV, we are sensitive to the running of
$\alpha_s(\mu)$ and are, therefore, able to extract values of 
$\Lambda_{\overline{\mathrm{MS}}}^{(5)}$.
We obtain
$\Lambda_{\overline{\mathrm{MS}}}^{(5)}=88{+34\atop-31}{+3\atop-23}$~MeV at
LO and
$\Lambda_{\overline{\mathrm{MS}}}^{(5)}=213{+75\atop-73}{+22\atop-29}$~MeV at
NLO, where the first errors are experimental and the second ones are
theoretical.
The experimental errors are determined by varying
$\Lambda_{\overline{\mathrm{MS}}}^{(5)}$ in such a way that the total
$\chi^2_{\mathrm{DF}}$ value is increased by one unit if all the other fit
parameters are kept fixed, while the theoretical errors are obtained by
repeating the LO and NLO fits for the scale choices $\xi=1/2$ and 2.
From the LO and NLO formulas for $\alpha_s^{(n_f)}(\mu)$, we thus obtain
\begin{eqnarray}
\alpha_s^{(5)}(M_Z)&=&0.1181{+0.0058\atop-0.0069}{+0.0006\atop-0.0049}\qquad
\mbox{(LO)},\nonumber\\
\alpha_s^{(5)}(M_Z)&=&0.1170{+0.0055\atop-0.0069}{+0.0017\atop-0.0025}\qquad
\mbox{(NLO)},
\end{eqnarray}
respectively.
As expected, the theoretical error is significantly reduced as we pass from LO
to NLO.
Adding the maximum experimental and theoretical deviations from the central 
values in quadrature, we find
$\Lambda_{\overline{\mathrm{MS}}}^{(5)}=(88\pm41)$~MeV and
$\alpha_s^{(5)}(M_Z)=0.1181\pm0.0085$ at LO and
$\Lambda_{\overline{\mathrm{MS}}}^{(5)}=(213\pm79)$~MeV and
$\alpha_s^{(5)}(M_Z)=0.1170\pm0.0073$ at NLO.
We observe that our LO and NLO values of $\alpha_s^{(5)}(M_Z)$ are quite
consistent with each other, which indicates that our analysis is 
perturbatively stable.
The fact that the respective values of
$\Lambda_{\overline{\mathrm{MS}}}^{(5)}$ significantly differ is a well-known
feature of the $\overline{\mathrm{MS}}$ definition of
$\alpha_s^{(n_f)}(\mu)$ \cite{cks}.

\begin{figure}[ht]
\begin{center}
\centerline{\epsfig{figure=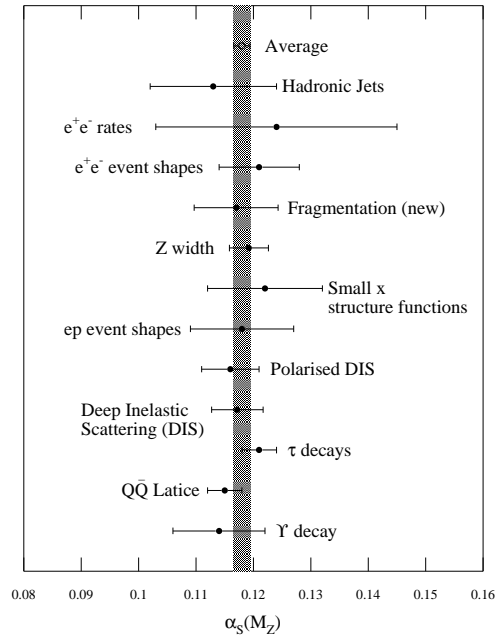,width=8cm}}
\caption{Summary of the values of $\alpha_s^{(5)}(M_Z)$ from various
processes.
The errors shown represent the total errors including theoretical
uncertainties.}  
\label{fig:two}
\end{center}
\end{figure}

Our values of $\Lambda_{\overline{\mathrm{MS}}}^{(5)}$ and
$\alpha_s^{(5)}(M_Z)$ perfectly agree with those presently quoted by the
Particle Data Group (PDG) \cite{pdg} as world averages,
$\Lambda_{\overline{\mathrm{MS}}}^{(5)}=208{+25\atop-23}$~MeV and
$\alpha_s^{(5)}(M_Z)=0.1181\pm0.002$, respectively.
Notice that, in contrast to our LO and NLO analyses, the PDG evaluates
$\Lambda_{\overline{\mathrm{MS}}}^{(5)}$ from $\alpha_s^{(5)}(M_Z)$ using the
three-loop relationship \cite{cks}.
The PDG combines twelve different kinds of $\alpha_s^{(5)}(M_Z)$ measurements, 
including one from the scaling violations in the FFs \cite{bus}, by
minimizing the total $\chi^2$ value and thus obtains
$\alpha_s^{(5)}(M_Z)=0.1181\pm0.0014$ with $\chi^2=3.8$.
The world average cited above is then estimated from the outcome by allowing
for correlations between certain systematic errors.
It is interesting to investigate how the world average of
$\alpha_s^{(5)}(M_Z)$ is affected by our analysis.
If we replace the value $\alpha_s^{(5)}(M_Z)=0.125\pm0.005\pm0.008$ resulting
from previous FF analyses \cite{bus}, which enters the PDG average, with
our new NLO value, then we obtain $\alpha_s^{(5)}(M_Z)=0.1180\pm0.0014$ with
$\chi^2=3.22$, {\it i.e.}, the face value of the world average essentially
goes unchanged, while the overall agreement is appreciably improved.
This is also evident from the comparison of Fig.~\ref{fig:two}, which
summarizes our updated world average, with the corresponding Fig.~9.1 in
Ref.~\cite{pdg}.
We observe that the central value of our new NLO result for
$\alpha_s^{(5)}(M_Z)$ falls into the shaded band, which indicates the error
of the world average, while in Fig.~9.1 of Ref.~\cite{pdg} the corresponding 
central value \cite{bus} exceeds the world average by 3.5 standard deviations
of the latter, which is more than for all other eleven processes.
Furthermore, our new NLO result has a somewhat smaller error (0.0073) than the
corresponding result \cite{bus} used by the PDG (0.009).
This is due to a marked decrease in the theoretical error, which may be
attributed to a different choice of input data, especially at low CM energies.
If we take the point of view that our new NLO value of $\alpha_s^{(5)}(M_Z)$
should rather be combined with the result from the previous FF analyses 
\cite{bus} before taking the world average, then the latter turns out to be
$\alpha_s^{(5)}(M_Z)=0.1181\pm0.0014$ with $\chi^2=3.34$.

\section{Global Analysis of Collider Data
\label{sec:four}}

We now review recent tests of the scaling violations and the universality of
our FFs \cite{kkp2}.
On the one hand, we confronted new data on inclusive charged-hadron production
from LEP2 \cite{D2}, with CM energies ranging from 133~GeV to 189~GeV, with
our NLO predictions.
On the other hand, we performed a global NLO analysis of essentially all 
high-statistics data on inclusive charged-hadron production in colliding-beam
experiments, including $p\bar p$ scattering at $Sp\bar pS$ \cite{UA1} and the
Tevatron \cite{CDF}, $\gamma p$ scattering at HERA \cite{H1}, and
$\gamma\gamma$ scattering at LEP2 \cite{Ogg}.
In the cases of hadroproduction and photoproduction, we set $\mu=M_f=\xi p_T$.
As for the parton density functions (PDFs) of the proton, we employ set CTEQ5M
provided by the CTEQ Collaboration \cite{CTEQ5}, with
$\Lambda_{\overline{\mathrm{MS}}}^{(5)}=226$~MeV.
As for the photon PDFs, we use the set by Aurenche, Fontannaz, and Guillet 
(AFG) \cite{AFG}.
In the following, we always consider the sum of positively and negatively 
charged hadrons.

\begin{figure}[ht]
\begin{center}
\begin{tabular}{ll}
\parbox{8cm}{\epsfig{figure=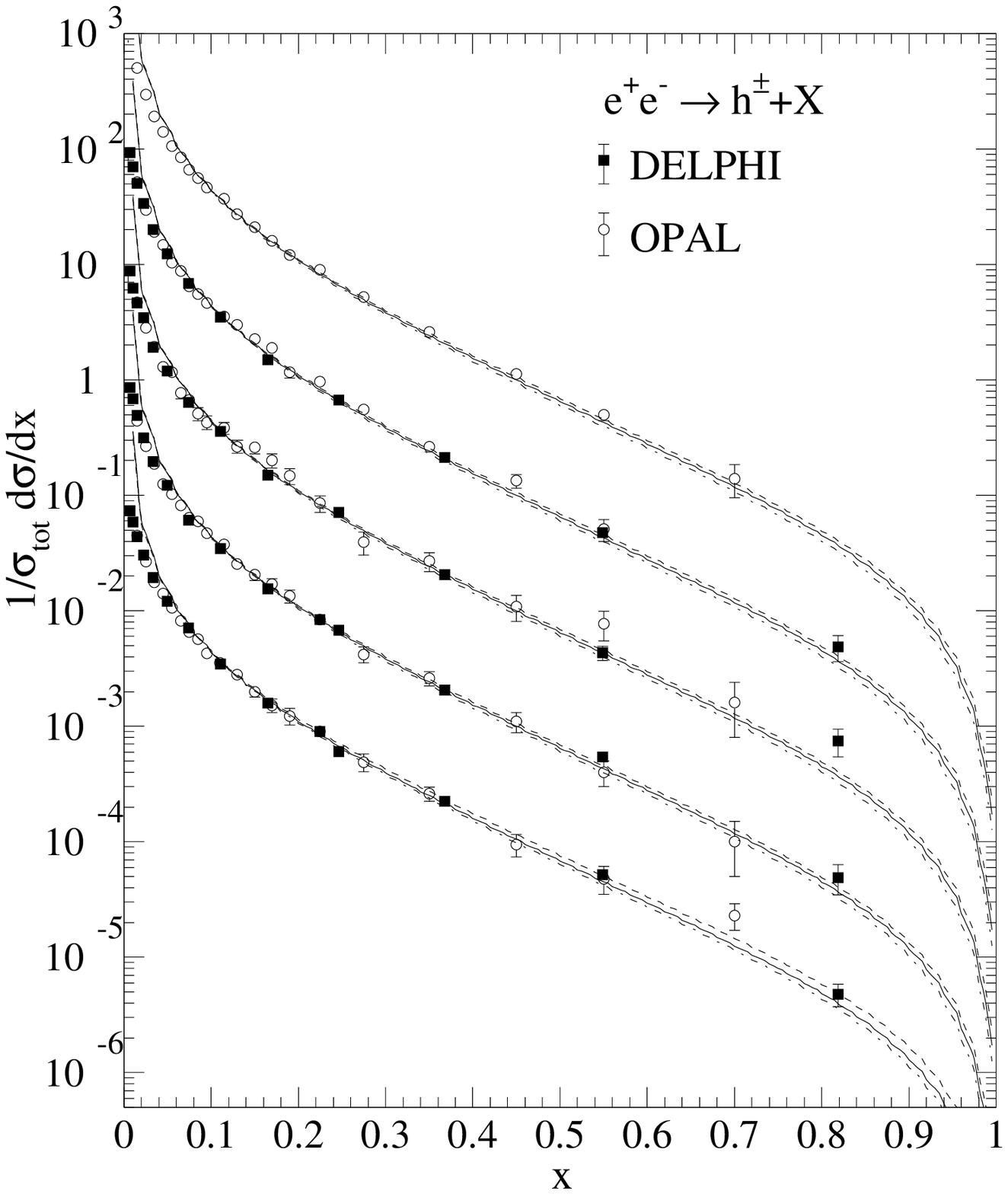,height=8cm}} &
\parbox{8cm}{\epsfig{figure=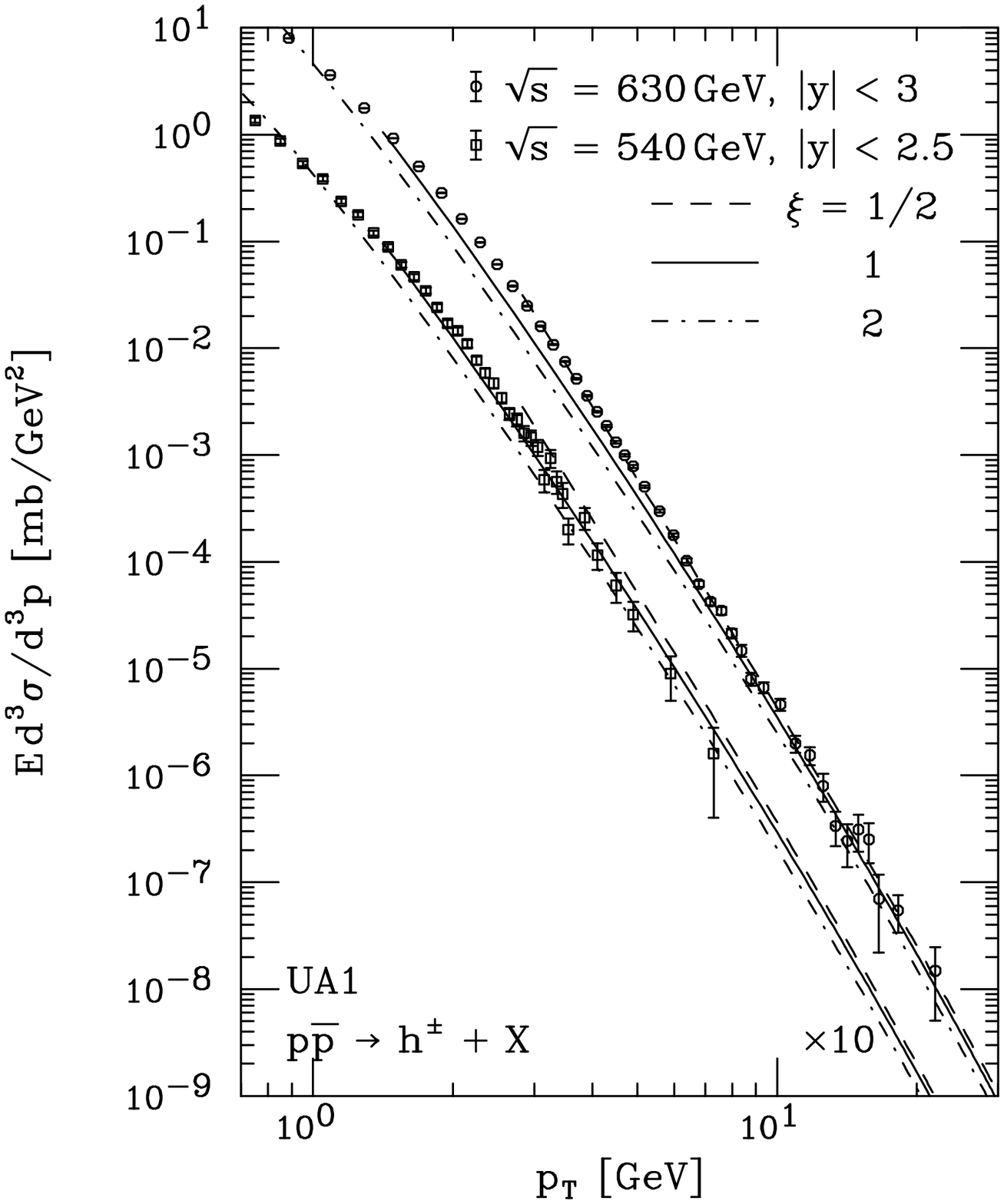,height=8cm}} \\
(a) & (b) \\
\parbox{8cm}{\epsfig{figure=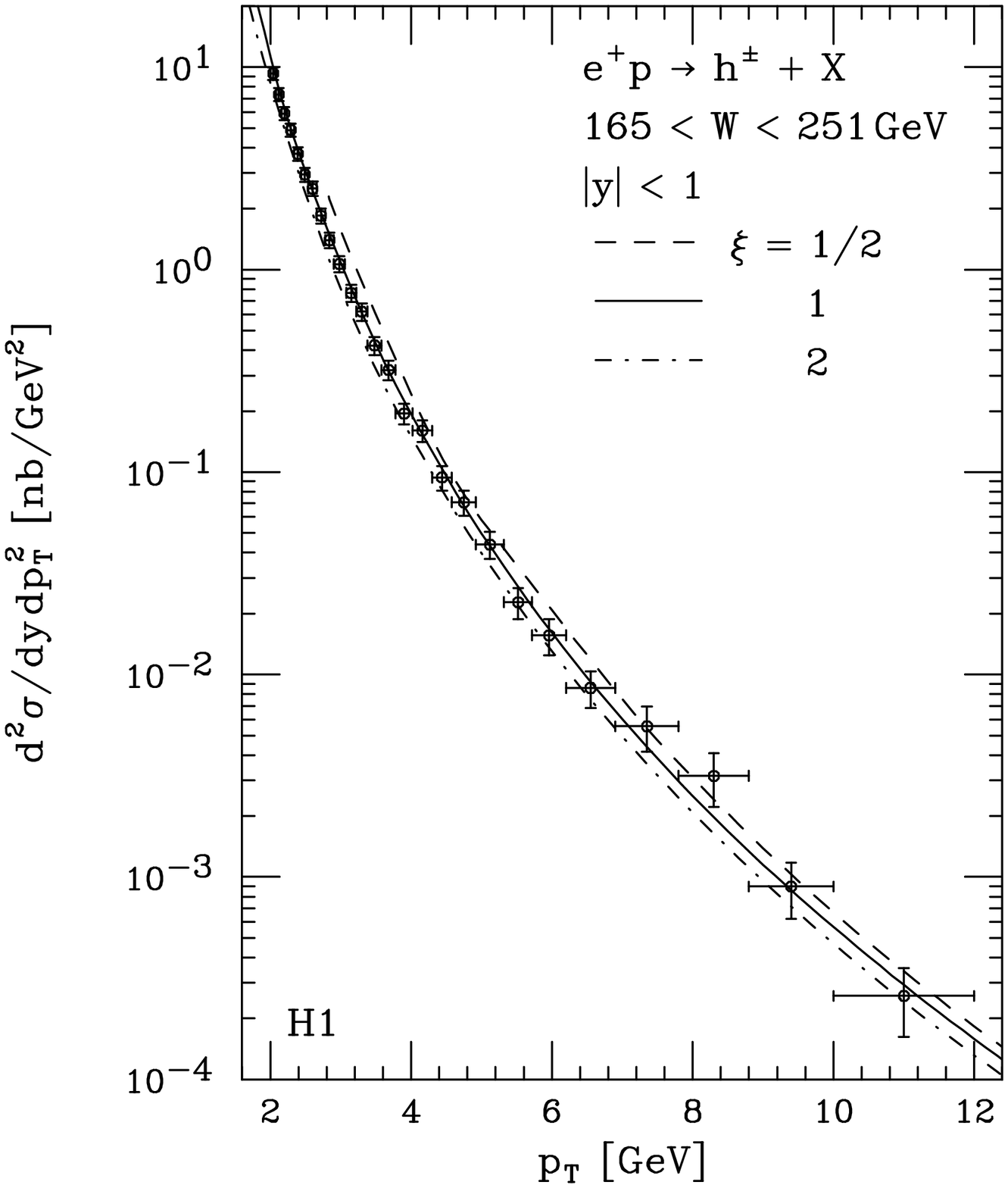,height=8cm}} &
\parbox{8cm}{\epsfig{figure=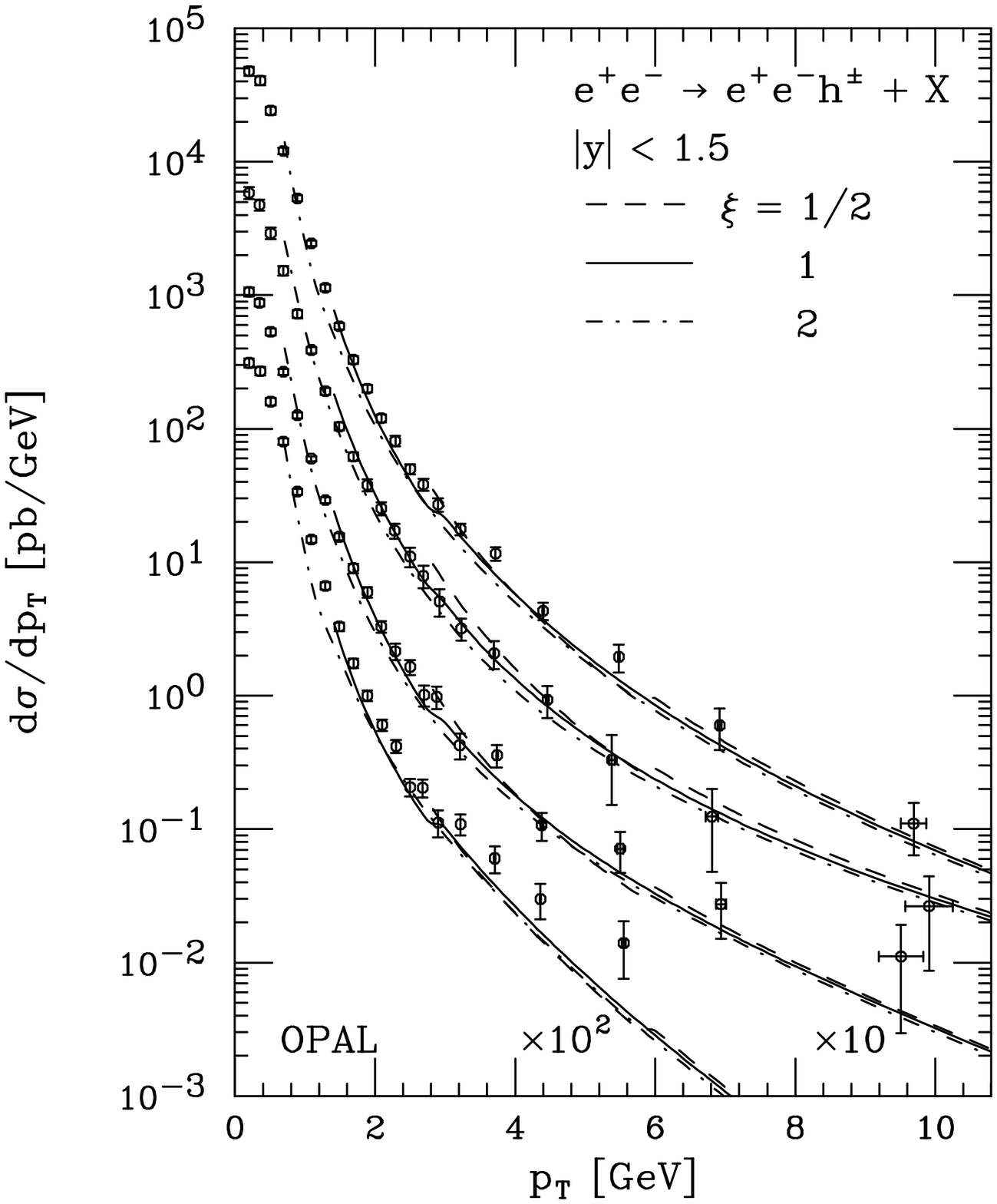,height=8cm}} \\
(c) & (d)
\end{tabular}
\caption{Comparisons of data on inclusive charged-hadron production in (a)
$e^+e^-$ annihilation at LEP2 with $\protect\sqrt s=133$, 161, 172, 183, and
189~GeV (from bottom to top in this order), (b) $p\bar p$ scattering at
S$p\bar p$S, (c) $\gamma p$ scattering at HERA, and (d) $\gamma\gamma$
scattering at LEP2 integrated over $\gamma\gamma$-invariant-mass intervals
$10<W<30$~GeV, $30<W<55$~GeV, $55<W<125$~GeV, and $10<W<125$~GeV (from bottom
to top in this order) with NLO predictions based on our FFs.}
\label{fig:three}
\end{center}
\end{figure}

In all cases, we found reasonable agreement between the experimental data and 
our NLO predictions as for both normalization and shape, as may be seen from
Fig.~\ref{fig:three}.
The majority of the data sets are best described with the central scale choice
$\xi=1$.
Exceptions include the UA1 data sets with $\sqrt s=200$ and 630~GeV and the
UA2 data set with $1<|y|<1.8$, which prefer $\xi=1/2$, as well as the ZEUS
data, which favours $\xi=2$.
However, if we estimate the theoretical uncertainty due to unknown corrections
beyond NLO by varying $\xi$ between 1/2 and 2, as is usually done, then it is
justified to state that all the considered data sets agree with our NLO
predictions within their errors.
We hence conclude that our global analysis of inclusive charged-hadron 
production provides evidence that both the predicted scaling violations and
the universality of the FFs are realized in nature.

\section{Conclusions
\label{sec:five}}

Owing to the high-statistics experimental information on how partons fragment
into low-mass charged hadrons provided by the LEP1 and SLC experiments, the
determination of NLO FFs advanced to a level of precision which is comparable
to the one familiar from similar analyses for PDFs.
In this presentation, we reviewed recent LO and NLO analyses of $\pi^\pm$,
$K^\pm$, and $p/\bar p$ FFs \cite{kkp}, which also yielded new values for
$\alpha_s^{(5)}(M_Z)$ \cite{kkp1}.
Although these FFs are genuinely nonperturbative objects, they possess two
important properties that follow from perturbative considerations within
the QCD-improved parton model and are amenable to experimental tests, namely
scaling violations and universality.
The scaling violations were tested \cite{kkp,kkp2} by making comparisons with
data of $e^+e^-$ annihilation at CM energies below \cite{low} and above
\cite{D2} those pertaining to the data that entered the fits.
The universality property was checked \cite{kkp2} by performing a global study
of high-energy data on hadroproduction in $p\bar p$ collisions \cite{UA1,CDF}
and on photoproduction in $e^\pm p$ \cite{H1} and $e^+e^-$ \cite{Ogg}
collisions.

As is well known, the gluon FF enters the prediction for the unpolarized cross
section of inclusive hadron production in $e^+e^-$ annihilation only at NLO,
while at LO it only contributes indirectly via the $\mu^2$ evolution.
In order to nevertheless have a handle on it, we included in our fits
\cite{kkp} experimental data on gluon-tagged three-jet events from LEP1
\cite{Ag,Og}.
Furthermore, we checked that our predictions for the longitudinal cross
section, where it already enters at LO, agree well with available data
\cite{Al,Dl}.
On the other hand, the gluon FF is known to play a crucial r\^ole for
$p\bar p$, $\gamma p$, and $\gamma\gamma$ scattering at low values of $p_T$.
Thus, the comparisons performed here provide another nontrivial test of the
gluon FF.

As we have seen in Fig.~\ref{fig:three}, the theoretical uncertainty of the
NLO predictions due to scale variations significantly decreases as $p_T$
increases.
In order to perform more meaningful comparisons, it would, therefore, be
desirable if $p\bar p$, $\gamma p$, and $\gamma\gamma$ experiments extended
their measurements out to larger values of $p_T$.
Furthermore, in order to render such comparisons more specific, it would be
useful if these experiments provided us with separate data samples of
$\pi^\pm$, $K^\pm$, and $p/\bar p$ hadrons.

We also estimated the current systematic uncertainties in the FFs by comparing
our NLO predictions for flavour-tagged inclusive charged-hadron production
with those obtained from two other up-to-date FF sets \cite{kre,bou}.
Apart from a difference in the $b$-quark FFs at medium to large values of $x$,
which may be traced to the incompatibility of two underlying
$b$-quark-specific data samples \cite{D,D1}, all three FF sets 
\cite{kkp,kre,bou} mutually agree within the present experimental errors.

\vspace{1cm}
\noindent
{\bf Acknowledgements}
\smallskip

\noindent
The author thanks the organizers of the symposium {\it 50 Years of Electroweak
Physics} in honor of Professor Alberto Sirlin's 70th birthday for the
invitation and for creating such a stimulating atmosphere.
He is grateful to G. Kramer and B. P\"otter for their collaboration on the
work presented here.
His research is supported in part by the Deutsche Forschungsgemeinschaft
through Grant No.\ KN~365/1-1, by the Bundesministerium f\"ur Bildung und
Forschung through Grant No.\ 05~HT9GUA~3, and by the European Commission
through the Research Training Network {\it Quantum Chromodynamics and the Deep
Structure of Elementary Particles} under Contract No.\ ERBFMRX-CT98-0194.

\end{document}